\documentclass[twocolumn]{aastex7}

\usepackage{amsmath,bm}
\hypersetup{pdfauthor={Name}}

\begin{document}

\title{The Feasibility of a Spacecraft Flyby with the Third Interstellar Object 3I/ATLAS from Earth or Mars}

\author[0009-0001-9538-1971]{Atsuhiro Yaginuma} 
\affiliation{Dept. of Physics and Astronomy, Michigan State University, East Lansing, MI 48824, USA}
\email[show]{yaginuma@msu.edu}

\author[0009-0000-4697-5450]{Tessa Frincke}
\affiliation{Dept. of Physics and Astronomy, Michigan State University, East Lansing, MI 48824, USA}
\email{frincket@msu.edu}

\author[0000-0002-0726-6480]{Darryl Z. Seligman}
\altaffiliation{NSF Astronomy and Astrophysics Postdoctoral Fellow}
\affiliation{Dept. of Physics and Astronomy, Michigan State University, East Lansing, MI 48824, USA}
\email{dzs@msu.edu}

\author[0000-0001-8397-3315]{Kathleen Mandt}
\affiliation{NASA Goddard Space Flight Center, Greenbelt, MD, 20771, USA}
\email{kathleen.mandt@nasa.gov}

\author[0000-0002-5643-1956]{Daniella N. DellaGiustina}
\affiliation{Lunar and Planetary Laboratory, University of Arizona, Tucson, AZ, USA.}
\email{dellagiu@arizona.edu}

\author[0000-0002-7257-2150]{Eloy Pe\~na-Asensio}
\affiliation{Department of Aerospace Science and Technology, Politecnico di Milano, Via La Masa 34, 20156 Milano, Italy}
\email{eloy.pena@polimi.it}

\author[0000-0002-0140-4475]{Aster G. Taylor}
\altaffiliation{Fannie and John Hertz Foundation Fellow}
\affiliation{Dept. of Astronomy, University of Michigan, Ann Arbor, MI 48109, USA}
\email{agtaylor@umich.edu}

\author[0000-0001-8316-0680]{Michael C. Nolan}
\affiliation{Lunar and Planetary Laboratory, University of Arizona, Tucson, AZ, USA.}
\email{mcn1@arizona.edu}

\begin{abstract}
We investigate the feasibility of a spacecraft mission to conduct a flyby of 3I/ATLAS, the third macroscopic interstellar object discovered on July 1 2025, as it traverses the Solar System. There are both ready-to-launch spacecraft currently in storage on Earth, such as Janus, and spacecraft nearing the end of their missions at Mars. We calculate minimum $\Delta V$ single-impulse direct transfer trajectories to 3I/ATLAS both from Earth and from Mars. We consider launch dates spanning January 2025 through March 2026 to explore obtainable and hypothetical mission scenarios. Post-discovery Earth departures require a challenging $\Delta V\gtrsim24$ km s$^{-1}$ to fly by 3I/ATLAS. By contrast, Mars departures from July 2025 - September 2025 require $\Delta V\sim5$ km s$^{-1}$ to achieve an early October flyby --- which is more feasible with existing propulsion capabilities. \added{We further calculate the phase angle and flyby velocity for these trajectories, noting that the resulting flyby speeds would impose significant observational and engineering challenges that a mission would need to overcome.} We discuss how existing spacecraft could be used to observe 3I/ATLAS and how spacecraft at other locations in the Solar System could be repurposed to visit future interstellar objects on short notice.

\end{abstract}


\section{Introduction} \label{sec:intro}

\setcounter{footnote}{0}

The discovery of 3I/ATLAS \citep{Denneau2025} increases the interstellar object sample by $50\%$, building on the previous discoveries of 1I/`Oumuamua in 2017 \citep{Williams17} and 2I/Borisov in 2019 \citep{borisov_2I_cbet}. On 2025 July 1, the Asteroid Terrestrial-impact Last Alert System (ATLAS) survey \citep{Tonry2018a} discovered the small body A11pl3Z \footnote{\url{https://www.minorplanetcenter.net/mpec/K25/K25N12.html}}. Fits to the ATLAS astrometry revealed that this object has a perihelion distance $ q\sim 1.36$ au, orbital eccentricity of $e\sim 6.2$, inclination of $i\sim175^\circ$, and a hyperbolic velocity of $v_\infty\sim58$ km s$^{-1}$, clearly indicating an interstellar origin. This object is only the third confirmed macroscopic interstellar object and therefore presents an unprecedented opportunity to characterize a planetesimal from outside of the Solar System.

Preliminary observations of 3I/ATLAS revealed a weakly active nucleus \citep{Jewitt2025a, Alarcon2025,Chandler2025} with a reddened reflectance spectrum and an absolute magnitude of $H_V\sim12$ \citep{Seligman2025, Opitom2025} although the nucleus size is uncertain \citep{Loeb2025}. Its visible and near-infrared reflectance is red and featureless, matching the spectral slopes of D-type asteroid and active Centaurus \citep{Belyakov2025, Kareta2025, Marcos2025}, \added{consistent with \textit{Seimei/TriCCS} photometry showing $g-r = 0.60\pm0.03$ \citep{Beniyama2025}.} High angular resolution observations from the Hubble Space Telescope WFC3  provided an upper limit on the nucleus radius of $\lesssim2.8$ km \citep{Jewitt2025b}. \added{\textit{Southern Astrophysical Research Telescope (SOAR)} r'-band photometry obtained days after discovery showed mean magnitudes of 17.5--18.1 and no clear long-term variability \citep{Frincke2025}. Prediscovery data from the  \textit{Zwicky Transient Facility (ZTF)} show activity starting near 9 au, with dust production increasing from $\sim5$ to $\sim30$ kg s$^{-1}$ between 6 and 4 au \citep{Ye2025}, and ATLAS monitoring revealed a color transition from red, $c-o \sim 0.7$, to nearly solar, $c-o \sim 0.3$, with appearance of a tail \citep{Tonry2025}.} Optical and near-infrared spectra revealed a broad 2.0 $\mu$m band consistent with $\sim 30 \%$ 10 $\mu$m-sized water ice grains in the coma \citep{Yang2025}. \added{Observations with the \textit{Neil Gehrels-Swift Observatory} revealed the presence of H$_2$O outgassing via the detection of OH emission \citep{Xing2025}, and SPHEREx and JWST found strong CO$_2$ emission (CO$_2$/H$2$O $\sim$7.6) with a dust-dominated continuum \citep{Lisse2025, Cordiner2025} Ground-based spectra revealed the onset of CN emission with Q$_{CN}$$ \sim$4--8$ \times10^{23}$ s$^{-1}$, strong carbon-chain depletion, and numerous Ni I lines without Fe I, suggesting unusual release mechanisms \citep{SalazarManzano2025, Rahatgaonkar2025}. Polarimetry showed a deep $-2.7\%$ branch at phase angle $\sim7^\circ$, and HST revealed a sunward anti-tail from anisotropic sublimation \citep{Gray2025, Keto2025}.} Time series photometry provided evidence of a rotation period of 16.16$\pm$0.01 h and dust loss rates between 0.3 and 4.2 kg s$^{-1}$ \citep{Santana-Ros2025}, while shift stack TESS pre-discoveries imply low level activity at $\sim6.4$ au \citep{Feinstein2025, Martinez-Palomera2025}. SOAR/Goodman spectra reveal a red continuum dominated by refractory organics, with no detected of CN, C$_2$, or CO$^{+}$ at 4.4 au \citep{Puzia2025}. 3I/ATLAS's high excess velocity and Galactic velocity radiant imply an old kinematic age of 3--11 Gyr \citep{Taylor2025} and lower-metallically stars, \added{suggesting thick-disk kinematics but still compatible with a thin-disk source \citep{Eubanks2025, Guo2025, Perez-Couto2025}.}

\added{The first two interstellar objects exhibited remarkably different characteristics. 1I/`Oumuamua showed no cometary activity yet displayed a significant nongravitational acceleration \citep{Meech2017,Ye2017,Jewitt2017,Trilling2018,Micheli2018}, whereas 2I/Borisov was clearly cometary with a well-developed coma \citep{Jewitt2019b,Fitzsimmons:2019,Ye:2019,McKay2020,Guzik:2020,Hui2020,Kim2020,Cremonese2020,yang2021}. Both had lower excess velocities than 3I/ATLAS, implying younger kinematic ages though with large uncertainty \citep{Mamajek2017,Gaidos2017a,Feng2018,Fernandes2018,Hallatt2020,Hsieh2021}. This diversity highlights that each newly detected object can provide insight into a different region of parameter space and that generalizations about their origins remain premature. In this context, \textit{in situ} spacecraft flybys or rendezvous are particularly powerful because measurements of H$_2$O abundances and isotopic ratios can trace the stellar populations from which these bodies formed \citep{Lintott2022,Hopkins2023,Hopkins2025a}, while the carbon-to-oxygen ratio constrains the formation location within a protoplanetary disk \citep{Seligman2022}. When combined with determinations of refractory organics and other volatiles, such data directly connect interstellar bodies to models of planetesimal formation and the delivery of volatiles to planetary atmospheres. Since analogous bodies may enrich short-period exoplanets with water and organics \citep{Seligman2022b}, a flyby mission to 3I/ATLAS would provide a unique opportunity to link small-body compositions to galactic chemical evolution and extrasolar habitability \citep{Seligman2018,Moore2021}.}

Many spacecraft have conducted close-range remote observations of comets and asteroids in the Solar System, such as the Deep Impact \citep{AHearn2005Sci}, Giotto \citep{Reinhard1986, Reinhard1987}, NEAR Shoemaker \citep{Prockter2002AcAau}, Rosetta \citep{Glassmeier2007}, DART \citep{Cheng2023}, and Hera \citep{Michel2022} missions. Moreover, a number of sample-return missions have delivered extraterrestrial material back to Earth, e.g., Stardust \citep{Brownlee2014AREPS, Brownlee1996}, Hayabusa \citep{Yoshikawa2021}, Hayabusa2 \citep{Watanabe2017}, and OSIRIS-REx \citep{Lauretta2017}.

Recently, Comet Interceptor --- an ESA F-class mission --- has been approved and is currently in its implementation phase \citep{Snodgrass2019, Jones2024}. Designed as a flexible mission architecture, Comet Interceptor will be stationed at the Sun–Earth L2 point, awaiting the discovery of a dynamically new comet or other pristine small body to intercept. Unlike previous missions, Comet Interceptor will employ a multi-spacecraft configuration to enable simultaneous multi-point observations of the target. This mission's primary objective is to investigate a long-period comet entering the inner Solar System for the first time. However, if early warning conditions permit, it could be redirected to perform the first \textit{in situ} exploration of an interstellar object.

The discovery of 1I/`Oumuamua and 2I/Borisov rapidly spurred development on an interstellar object mission concept. Initial studies (e.g., \citealt{Seligman2018}) emphasized the urgency of preparing interception missions with minimal latency but determined that it was feasible to detect and reach future interstellar objects with conventional chemical propulsion. Further studies such as Project Lyra examined the feasibility of direct missions to 1I/`Oumuamua using near-term technologies, including solar Oberth maneuvers and high-energy launch architectures \citep{Hein2019, Hibberd2021A, Hibberd2023}. These analyses demonstrated that with optimized trajectories and sufficient launch energies, flyby or rendezvous missions could be viable even years after 1I/`Oumuamua’s perihelion. Alternative mission design proposals include solar sail architectures capable of prolonged loitering and rapid-response interception of future interstellar objects \citep{Hein2019, Garber2022, Miller2022}, or a distributed statite cluster for high-$\Delta V$ response scenarios \citep{Hoover2022}. Evaluations of optical and autonomous navigation performance during high-velocity flybys emphasize the importance of on-board image processing and neural network guidance to maintain targeting accuracy at encounter speeds exceeding 60 km s$^{-1}$ \citep{Mages2022, Donitz2023}. A complementary analysis focused on the tradeoffs between target detectability, spacecraft storage strategies, and imaging resolution advocated for rendezvous missions to 1I/`Oumuamua-sized targets \citep{Siraj2023}. \citet{Stern2024} also investigated the feasibility of an intercept mission to an interstellar object, ``Interstellar Object Explorer.'' The ``Bridge'' New Frontiers-class flyby mission concept would intercept an interstellar object with a dedicated science payload including a guided impactor, ultraviolet/visible/infrared spectrometers, and a high-resolution camera \citep{Moore2021}. Additional studies explored broader science cases and mission architectures for interstellar object investigation, including rapid-response trajectories and programmatic feasibility \citep{Donitz2021}, the astrobiological relevance of interstellar objects as potential carriers of organic material and the propulsion technologies that could enable their exploration \citep{Lingam2023}, and the practical constraints and feasibility of near-term rendezvous missions to interstellar objects using advanced but achievable propulsion architectures \citep{Landau2023}.

Here, we investigate possible missions that could reasonably intercept 3I/ATLAS in the months following its July 2025 discovery. We specifically explore utilizing existing missions near Earth or Mars and consider favorable minimum distances to 3I/ATLAS. This paper is structured as follows --- in Section \ref{sec:methods}, we detail an algorithm designed to determine direct transfer   trajectories and  launch dates. In Section \ref{sec:earth_mission}, we calculate minimum-energy mission trajectories to 3I/ATLAS from Earth achievable in the months following its discovery. Here, we defined minimum-energy as the trajectory requiring the lowest single‐impulse $\Delta V$. In Section \ref{sec:marsmission}, we identify additional key launch dates and trajectories for a flyby mission to 3I/ATLAS from Mars. Finally, we conclude in Section \ref{sec:discussion} with a summary of our results and a final discussion of our suggestions for immediate implementation of an intercept mission with 3I/ATLAS. \added{Our analysis provides a concrete feasibility assessment tailored to a specific known interstellar body and demonstrates how near-term assets could be leveraged in practice.}

\section{Identification of Optimal Trajectories}\label{sec:methods}

We compute minimum-energy transfers using the universal-variable formulation of Lambert’s problem \citet{Sukhanov1989, Bate1971, Vallado2001}, applied to Earth and a hyperbolic target by following the elliptical-orbit formulation of \citet{Leeghim2013} and the hyperbolic-target extension in Section 4 of \citet{Seligman2018}. \added{The families of transfers shown as functions of departure date and time of flights are equivalent to the conventional "pork-chop plots." For continuity with this work, we refer to them as minimum-energy trajectory maps while noting their equivalence to the standard pork-chop framework.}

We sample departure epochs $\tau_0$ over a 400-day window and flight times $\Delta\tau$ across a mission-relevant range. Under a patched-conic approximation, the departure planet's heliocentric state, we take the heliocentric state $\big(\vec{r}_{0,I}, \vec{v}_h\big)$ at $\tau_0$ and the target position $\vec{r}_T$ at $\tau_0 + \Delta\tau$ are taken from JPL Horizons. Each grid point $(\tau_0,\Delta\tau)$ is solved as a single universal-variable Lambert problem to obtain the required departure velocity $\vec{v}_{0,I}$ and its associated direct-flight cost $\Delta V$. \added{Each trajectory corresponds to a direct Lambert solution between the heliocentric state of the departure planet at epoch $\tau_0$ and that of 3I/ATLAS at $\tau_0 + \Delta\tau$. In this way, the transfers reflect the object’s actual three-dimensional hyperbolic path through the Solar System, and are not confined to intersections with the ecliptic plane.}

We work with the standard universal-variable parameter $z\equiv\frac{\chi^2}{a}$ with semimajor axis $a$ and universal variable $\chi$. The transfer geometry is set by the angle between the interceptor position $\vec{r}_I\equiv\vec{r}_{0,I}$ and the known target position $\vec{r}_T$. We compute the angle between the target and interceptor, $\Delta\theta$, given by:

\begin{equation}\label{eq:geometry}
  \begin{split}
    \cos\big(\Delta\theta\big) &= \frac{\vec{r}_{T}\cdot\!\vec{r}_{I}}{\lVert\vec{r}_{T}\rVert\,\lVert\vec{r}_I\rVert}\,.
  \end{split}
\end{equation}

We define a geometric coefficient:

\begin{equation}\label{eq:geometry_para}
  \begin{split}
    A &= \sin\big(\Delta\theta\big)
         \,\sqrt{\frac{\lVert\vec{r}_{T}\rVert\,\lVert\vec{r}_I\rVert}{1-\cos\big(\Delta\theta\big)}}\,,
  \end{split}
\end{equation}

which rescales the chord connecting the departure and arrival position vectors so that the universal variable time-of-flight equation properly reflects the transfer geometry.

We then introduce the intermediate function:

\begin{equation}\label{eq:y_def}
  \begin{split}
    y(z) &= \lVert\vec{r}_{T}\rVert + \lVert\vec{r}_{I}\rVert
    + A\,\frac{z\,S(z) - 1}{\sqrt{C(z)}}\,.
  \end{split}
\end{equation}

Here, $S(z)$ and $C(z)$ denote the Stumpff functions in their conventional definitions (maybe cite).

\begin{equation}\label{eq:StumpffS}
  \begin{split}
    S(x) &\equiv \begin{cases}
    \displaystyle
    \frac{\sqrt{x} - \sin\!\bigl(\sqrt{x}\bigr)}{\bigl(\sqrt{x}\bigr)^3},
    & x>0\ (\text{elliptical}),\\
    \displaystyle
    \frac{1}{6},
    & x=0\ (\text{parabolic}),\\
    \displaystyle
    \frac{\sinh\!\bigl(\sqrt{-x}\bigr) - \sqrt{-x}}{\bigl(\sqrt{-x}\bigr)^3},
    & x<0\ (\text{hyperbolic}), \end{cases}
  \end{split}
\end{equation}

and

\begin{equation}\label{eq:StumpffC}
    \begin{split}
        C(x)&=\begin{cases}
        \displaystyle
        \frac{1 - \cos\!\bigl(\sqrt{x}\bigr)}{x},
        & x>0\quad(\text{elliptical}),\\
        \displaystyle
        \frac{1}{2},
        & x=0\quad(\text{parabolic}),\\
        \displaystyle
        \frac{\cosh\!\bigl(\sqrt{-x}\bigr) - 1}{-\,x},
        & x<0\quad(\text{hyperbolic}). \end{cases}
    \end{split}
\end{equation}

With Equation \ref{eq:y_def}, the time-of-flight residual can be written as defined in \citep[Chapter 5, Eq. 5.40]{Curtis2020}:

\begin{equation}\label{eq:F_def}
  \begin{split}
    F(z) 
    &= \Bigl(\frac{y(z)}{C(z)}\Bigr)^{3/2}\!S(z)
      + A\,\sqrt{y(z)}
      - \sqrt{{GM_\odot}}\,\Delta\tau\,,
  \end{split}
\end{equation}

Root-finding on Equation \ref{eq:F_def} enforces both geometric and temporal constrains simultaneously. For a root $z$, the Lambert-Lagrange coefficients are

\begin{equation}\label{eq:fg_coeffs}
  \begin{split}
    f_I &= 1 - \frac{y(z)}{\lVert\vec{r}_{0,I}\rVert},\\
    g_I &= A\,\frac{\sqrt{y(z)}}{\sqrt{GM_\odot}}.
  \end{split}
\end{equation}

The required arrival position and velocity vectors can be written as:

\begin{equation}\label{eq:rt}
    \vec{r}_{T} = f_I \vec{r}_{0,I} + g_I \vec{v}_{0,I},
\end{equation}
and
\begin{equation}\label{eq:voi}
  \begin{split}
    \vec{v}_{0,I} &= \frac{\vec{r}_{T}-f_I \vec{r}_{0,I}}{g_I}\,.
  \end{split}
\end{equation}

To evaluate the impulsive change in velocity $\Delta V$ required for a spacecraft to reach 3I/ATLAS, we subtract the heliocentric velocity of the departure planet from $\vec{v}_{0,I}$, so that:

\begin{equation}\label{eq:deltav}
  \begin{split}
    \Delta V &= \lVert \vec{v}_{0,I} - \vec{v}_{h} \bigr\rVert\,.
  \end{split}
\end{equation}

In Equation \ref{eq:deltav}, $\vec{v}_h$ is the heliocentric velocity of departure planet.

We repeat the following process iteratively to solve for $\Delta V$:

\begin{enumerate}
    \item  Choose a trial departure epoch $\tau_{0,I}$ and set a flight time $\Delta\tau$.
    \item  Solve the universal variable Lambert problem for that $\Delta\tau$ by finding the root $z$ of the time-of-flight residual defined in Equation \ref{eq:F_def}.
    \item Compute the Lagrange coefficient using Equation \ref{eq:fg_coeffs} and propagate the interceptor state with Equations \ref{eq:rt} and \ref{eq:voi}.
    \item Evaluate $\Delta V$ using Equation \ref{eq:deltav}.
    \item Identify the global minimum-energy trajectory from the resulting $\Delta V$ values.
\end{enumerate}

It is important to note that these $\Delta V$ values only encapsulate the hyperbolic-excess component of the impulsive burn required to leave a circular low-Earth orbit on an escape trajectory. In other words, they do not include the velocity needed to climb out of the departure planet's gravitational potential nor any downstream deep space maneuvers. \added{For each launch, we also report the characteristic energy C$_3$, which is the standard metric used to describe launch vehicle capability. In our formulation, $\Delta V$ corresponds to the hyperbolic excess velocity relative to Earth, so C$_3$ is simply given by $C_3 = v_\infty^2 = (\Delta V)^2$.}

\added{In addition to $\Delta V$ and $C_3$, we also report the flyby speed and the phase angle (see table \ref{table:trajectories}). The flyby speed is the spacecraft’s three-dimensional speed relative to 3I/ATLAS at flyby. The phase angle is the Sun–target–spacecraft angle at flyby, with the vertex at the target.}

Finally, we verified that this methodology accurately produced flyby trajectories. We performed numerical integrations with hypothetical massless test particles representing the spacecraft and the Sun-Earth system with the solution velocity vectors. We verified that the position of the spacecraft and target matched after the corresponding time of flight. All verification numerical simulations were performed with \texttt{REBOUND} N-body code \citep{rebound}, the \texttt{REBOUNDx} library \citep{reboundx} and \texttt{MERCURIUS} \citep{mercurius}.

The universal variable Lambert solver described above is a well-established solution to a two‐body problem \citep{Seligman2018, Curtis2020}. Our novel contribution is not in the solver itself but in its systematic application to the evolving geometry of 3I/ATLAS and subsequent mapping of the resulting minimum $\Delta V$ to the Earth- and Mars-based launchers. This enables the identification of mission-ready intercept windows under realistic departure and arrival constraints.

\begin{figure}
\centering
\includegraphics[width=1.\linewidth]{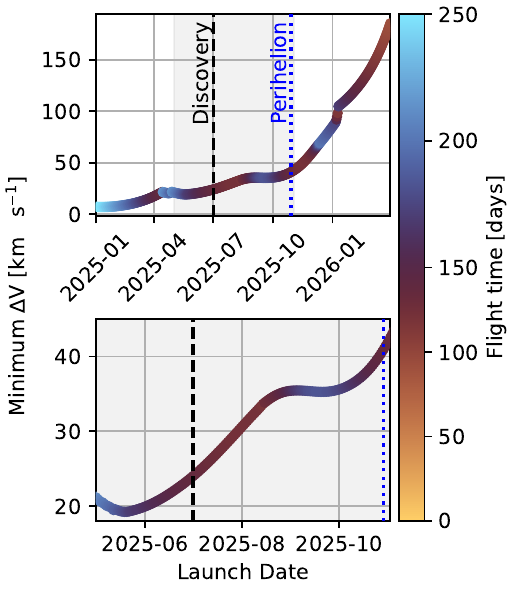}
\caption{
Direct flight $\Delta V$ values as a function of launch date for trajectories  to 3I/ATLAS from Earth.  The color of each point corresponds to the flight time of the mission. The upper- and lower-panels show direct flight $\Delta V$ values from 2025 January 1 -- 2026 March 31 and May 1 -- November 2 2025, respectively. The minimum post-discovery $\Delta V$ trajectory is 24.0 km s$^{-1}$ \added{(C$_3$ = 576 km$^{2}$ s$^{-2}$)} on 2025 July 1 with a flight time of 137 days. Pre-discovery trajectories may have been feasible, requiring $\Delta V\sim7$ km s$^{-1}$ with a flight times of $\sim$250 days. The shaded region on the upper panel is the region shown in the lower panel.
} 
\label{fig:earthmission_deltav}
\end{figure}

\begin{figure}
\centering
\includegraphics[width=1.\linewidth]{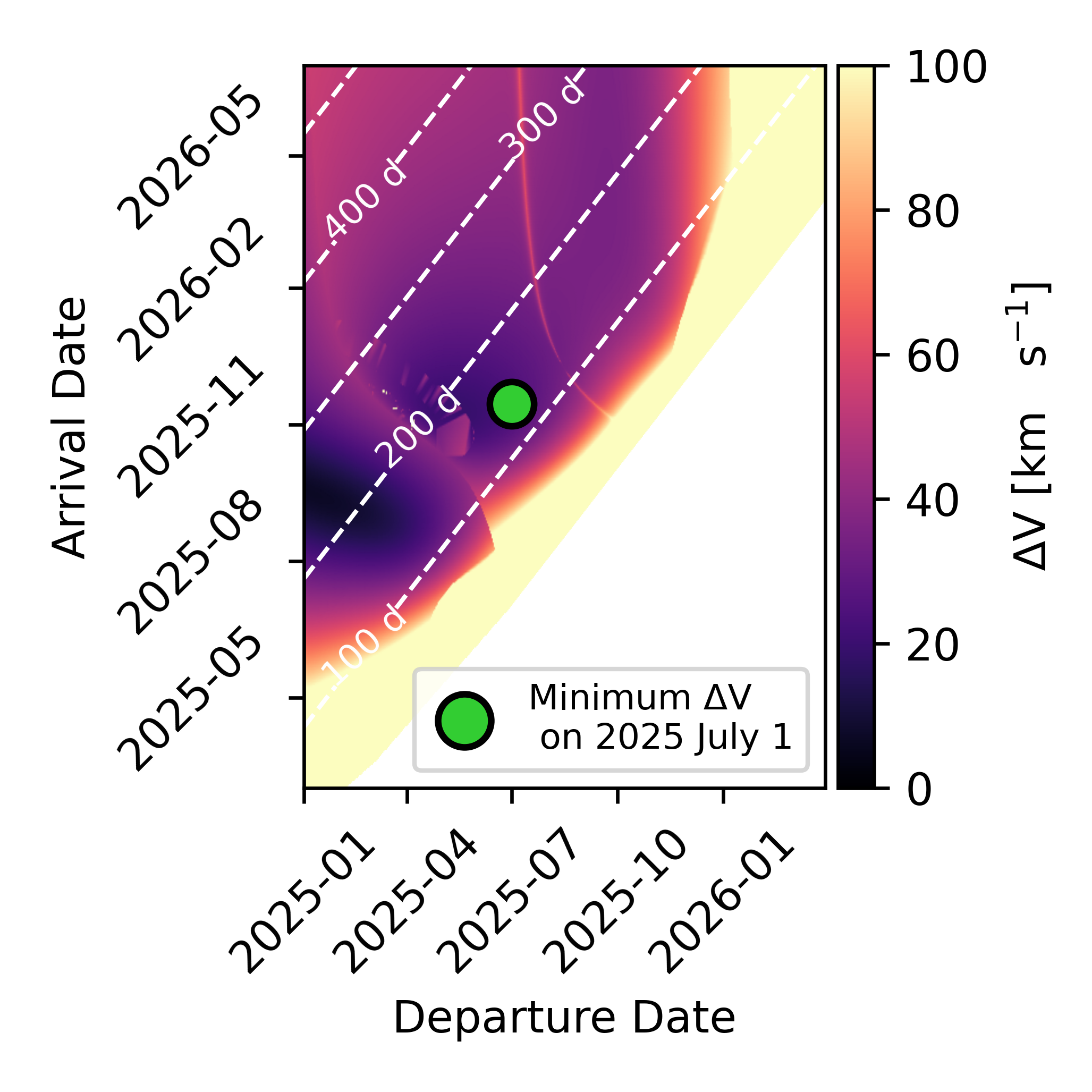}
\caption{
Mission Design Contour displaying required $\Delta V$ (0--100 km $s^{-1}$) across Earth departure and 3I/ATLAS arrival dates. The Green circle mark indicates the minimum $\Delta V$ on discovery date, 2025 July 1.
} 
\label{fig:earthmission_map}
\end{figure}

\section{Optimal Trajectories from Earth} \label{sec:earth_mission}

\begin{table*}
\caption{Velocity vector information of minimum-energy trajectories to 3I/ATLAS for a variety of launch dates and locations. All listed velocities are with respect to the reference frame comoving with the orbit of the Earth. }
\hspace*{-3cm}
\begin{tabular}{@{}llllllllll@{}}
\hline
Launch Date & Launch & $V_x$&$V_y$&$V_z$&$\Delta V$&Flight time&Arrival Date&Flyby Speed&Phase\\
 & Location & [km s$^{-1}$]&[km s$^{-1}$]&[km s$^{-1}$]&[km s$^{-1}$]&[days]& &[km s$^{-1}$]&Angle [$^\circ$]\\
\hline
2025 January 10 & Earth & -5.353 & \phantom{-}0.592 & \phantom{-}4.371 & \phantom{-}6.935 & 248 & 2025 September 15 & 79.96 & 49.0\\
2025 July 1 & Earth & -8.196 & \phantom{-}22.536 & -0.996 & \phantom{-}24.001 & 137 & 2025 November 15 & 79.73 & 95.4 \\
2025 December 15 & Earth & \phantom{-}8.424 & \phantom{-}70.532 & -4.082 & \phantom{-}71.151 & 198 & 2026 July 1 & 15.84 & 62.6 \\
2025 March 6 & Mars & -0.483 & -0.082 & \phantom{-}1.958 & \phantom{-}2.019 & 212 & 2025 October 4 & 86.19 & 66.8\\
2025 July 1 & Mars & -1.641 & -1.033 & \phantom{-}2.959 & \phantom{-}3.538 & 94 & 2025 October 3 & 86.43 & 65.2\\
2025 August 10 & Mars & -3.291 & -1.994 & \phantom{-}4.8354 & \phantom{-}6.179 & 54 & 2025 October 3 & 86.32 & 64.3 \\
2025 November 10 & Mars & \phantom{-}14.410 & \phantom{-}72.571 & \phantom{-}3.959 & \phantom{-}74.093 & 233 & 2026 July 1 & 18.88 & 46.9\\
\hline
\end{tabular}
\label{table:trajectories}
\end{table*}

\begin{figure*}
\centering
\includegraphics[width=1.\linewidth]{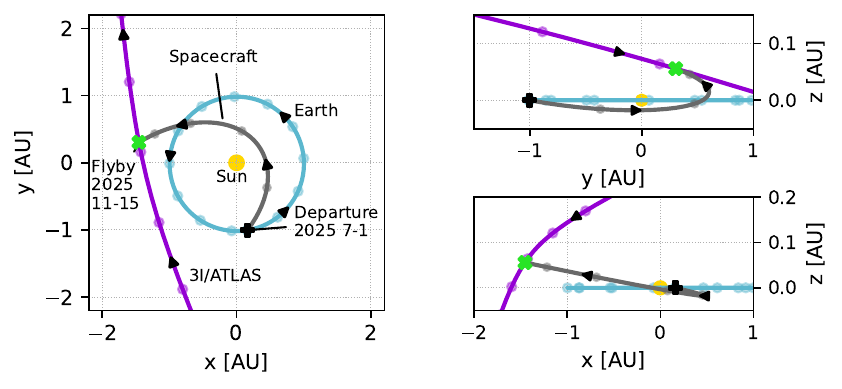}
\caption{The orbit of a post-discovery minimum $\Delta V$ mission to 3I/ATLAS  sent on 2025 July 1 from the Earth. The spacecraft would encounter 3I/ATLAS on 2025 November 15 and would  require $\Delta V= 24$ km s$^{-1}$. The trajectory for 3I/ATLAS, Earth, and the spacecraft are plotted in purple, blue, and gray, respectively. Markers represent departure location and flyby location, and arrows show direction of each orbit/trajectory. Points on each trajectory indicate the positions of each object every 30 days.}
\label{fig:earth_mission_orbit}
\end{figure*}

In this section, we examine the feasibility of trajectories from the Earth to the interstellar object 3I/ATLAS. We calculate minimum-energy trajectories from Earth to 3I/ATLAS using the methodology described in Section \ref{sec:methods}.

To determine when a direct launch from Earth could be viable, we use the Earth's position as the initial launch date for a spacecraft. We then calculate the minimum $\Delta V$ required to reach 3I/ATLAS for a range of possible initial launch dates. The minimum $\Delta V$ required as a function of launch date for a direct flight is shown in Figure \ref{fig:earthmission_deltav}. The color of each point indicates the flight time for each of the corresponding trajectories. Figure \ref{fig:earthmission_map} shows the required $\Delta V$ across Earth departure and 3I/ATLAS arrival date.

The minimum $\Delta V$ solution recovered after the discovery of the object technically occurs on 2025 July 1. In other words, the \added{minimum-energy trajectory} would have required immediate launch of a spacecraft contemporaneous with discovery. This trajectory would have required $\Delta V\sim 24.0$ km s$^{-1}$ \added{(C$_3$ = 576 km$^{2}$ s$^{-2}$)}, with a flight time of 137 days. The details of this trajectory are reported in Table \ref{table:trajectories}. An orbit schematic of this hypothetical minimum-energy flyby mission is shown in Figure \ref{fig:earth_mission_orbit}.

\added{Unlike typical interplanetary missions where both nodes are available to reduce out-of-plane cost, only one node is accessible during the short observational window for 3I/ATLAS. We also calculated a trajectory that flyby at the node, which requires $\Delta V \sim 26$ km s$^{-1}$ because the spacecraft must travel farther in a shorter time to reach it. Thus, the $\Delta V \sim 24$ km s$^{-1}$ case with high inclination represents the minimum-energy solution, even though it involves a large out-of-plane component.}

\added{This $\Delta V$ requirements for an Earth-based flyby underscore the limited practicality of such a mission with current propulsion capabilities. This emphasizes that a substantially earlier discovery would have been necessary for an intercept trajectory to fall within a realistically achievable range.}

The energetic requirements for a direct transfer mission post-discovery gradually increase during the timespan between July and early October 2025. Specifically the $\Delta V$ value increases to $\sim35$ km s$^{-1}$ while the flight time extends to $\sim140$ days. After early October 2025 --- but still before perihelion --- the minimum $\Delta V$ rises steeply. The minimum $\Delta V$ reaches a value of $\sim60$ km s$^{-1}$ by December 2025 and $\sim80$ km s$^{-1}$ in January 2026. Therefore, an Earth-based direct flight is significantly more attainable if a spacecraft is launched before 3I/ATLAS passes its perihelion. However, even $\Delta V = 24$ km s$^{-1}$ for a fiducial best case mission shown in Figure \ref{fig:earth_mission_orbit} would be challenging for current propulsion capabilities.

A direct-transfer trajectory from Earth would have been more energetically feasible if 3I/ATLAS had been discovered earlier. Earth departures between January and June would have required $\Delta V \sim 10$ km s$^{-1}$, with a flight time of approximately 150--200 days. A launch on 2025 January 10 would require only a minimum $\Delta V$ of 6.94 km s$^{-1}$ \added{(C$_3$ = 48.2 km$^{2}$ s$^{-2}$)} with a flight time of 248 days, which may have been within the reach of modern spacecraft. \citet{Sanchez2025} demonstrated that a complementary deep‐space maneuver (DSM) study for the European Space Agency Comet Interceptor mission (CI) could intercept 3I/ATLAS on 2025 November 7 after an $\sim 825$ days of flight time. This would require departing the Sun-Earth L$_2$ halo orbit on 2023 August 2 with a hyperbolic departure velocity of 2.8 km s$^{-1}$ and performing a 685 m s$^{-1}$ midcourse burn. It is worth noting that in a recent and complimentary analyses, \citet{Hibberd2025} identified an optimal trajectory from Earth to 3I/ATLAS. They calculated a minimum energy trajectory launched in 2024 July 19 for a rendezvous 2025 October 18. That trajectory assumes a single impulsive $\Delta V$ burn \added{with a C$_3$ requirement of 19.3 km$^{2}$ s$^{-2}$.}

\begin{figure}
\centering
\includegraphics[width=1.\linewidth]{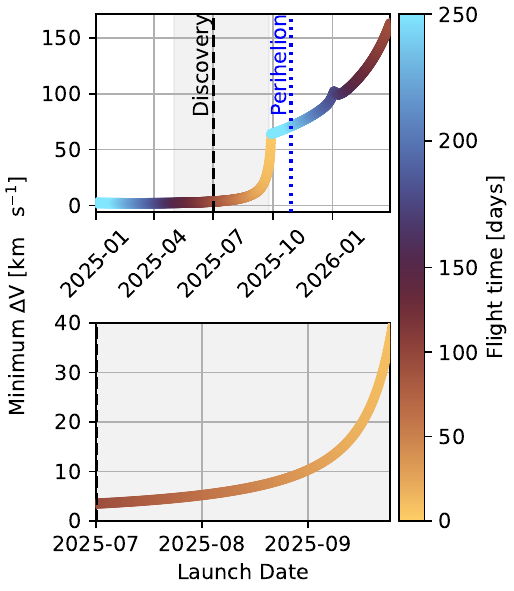}
\caption{Similar to Figure \ref{fig:earthmission_deltav}, but for trajectories to 3I/ATLAS from Mars. Here, the shaded inset region is from 2025 July 1 to 2025 September 25. The minimum energy trajectory for a post-discovery mission would be launched on 2025 July 1 with a flight time of 94 days and require $\Delta V=$  3.54 km s$^{-1}$. Pre-discovery trajectories would have required $\Delta V\sim3$ km s$^{-1}$ \added{(C$_3$ = 12.5 km$^{2}$ s$^{-2}$)} with flight times of more than 200 days.} 
\label{fig:mars_mission_deltav}
\end{figure}

\begin{figure}
\centering
\includegraphics[width=1.\linewidth]{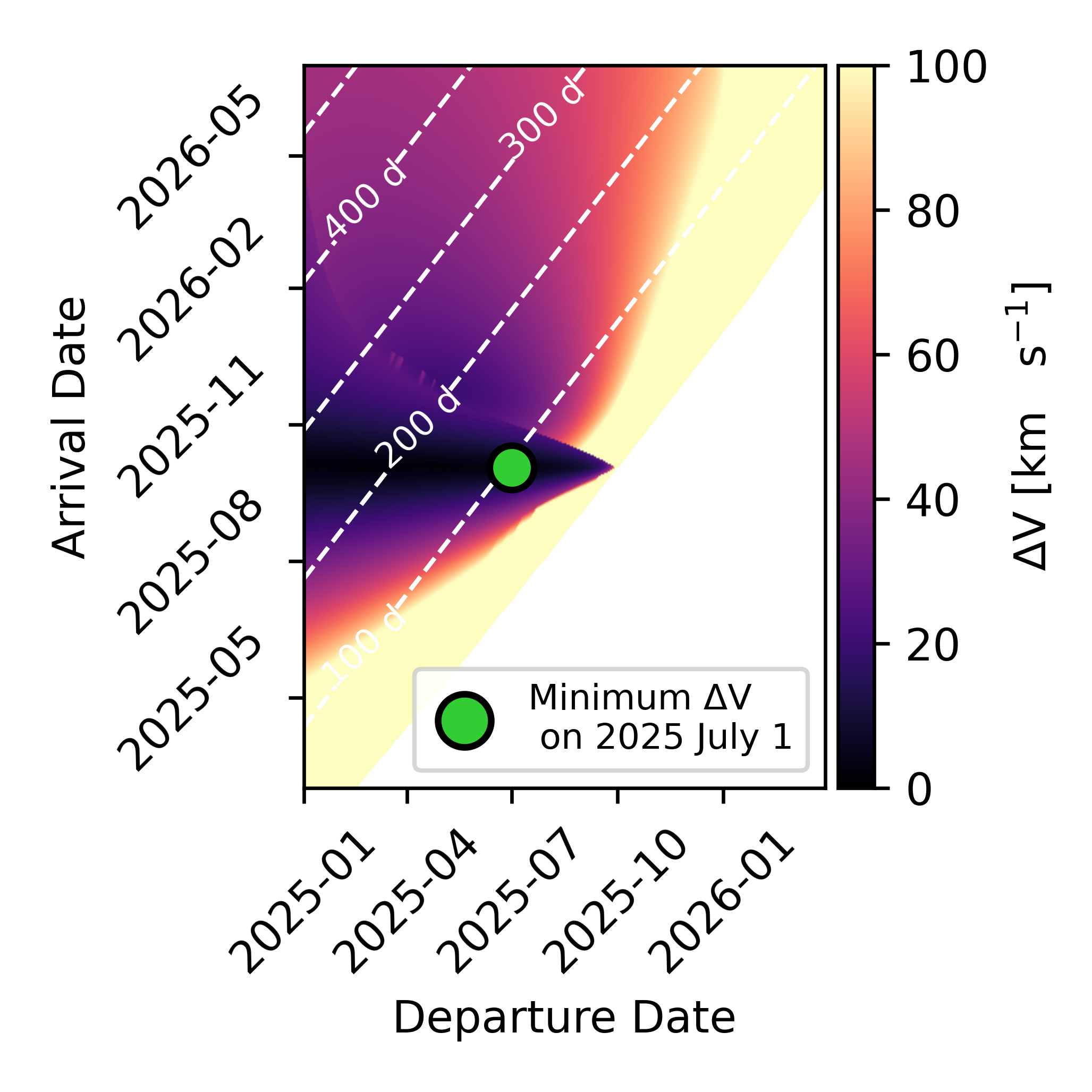}
\caption{
Similar to Figure \ref{fig:earthmission_map}, but from Mars in this case.
} 
\label{fig:marsmission_map}
\end{figure}

\begin{figure*}
\centering
\includegraphics[width=1.\linewidth]{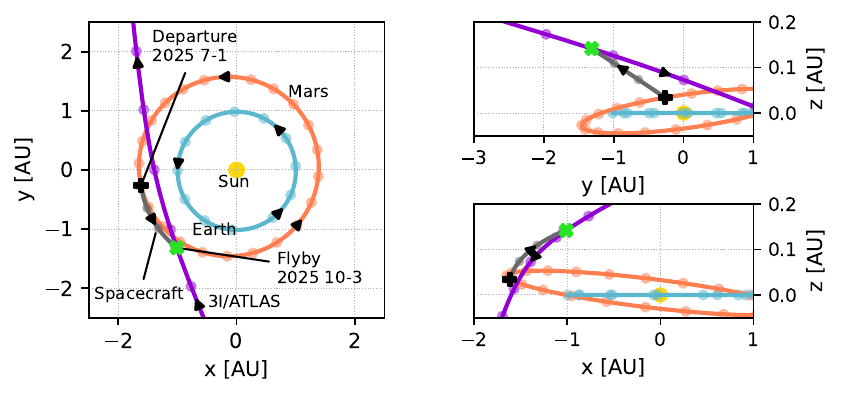}
\caption{Similar to Figure \ref{fig:earth_mission_orbit}, except for a minimum $\Delta$V mission to 3I/ATLAS post-discovery, sent on 2025 July 1 from Mars. This hypothetical trajectory requires a $\Delta V \sim 4$ km s$^{-1}$ and a time of flight of 94 days.}
\label{fig:mars_mission_orbit}
\end{figure*}

\begin{figure*}
\centering
\includegraphics[width=1.\linewidth]{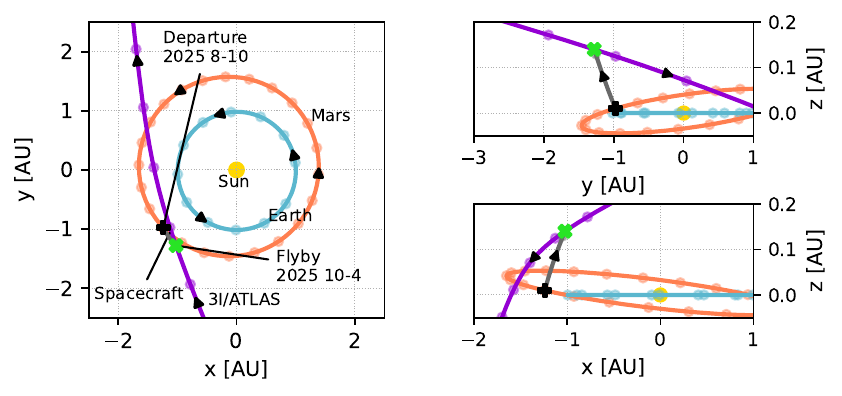}
\caption{Similar to Figure \ref{fig:earth_mission_orbit}, except for a minimum-energy mission to 3I/ATLAS sent from Mars on 2025 August 10. Such a mission would require a $\Delta V \sim 6.5$ km s$^{-1}$ and performs a flyby on 2025 October 4 with a time of flight of 54 days.}
\label{fig:mars_mission_orbit_2}
\end{figure*}

\section{Optimal Trajectories from Mars } \label{sec:marsmission}

Mars has been a high-priority planetary science target for multiple space agencies, including NASA and the ESA, for several decades. A typical mission starts with the ``prime mission'', which is designed to achieve the originally-funded objectives. Once a prime mission is completed, if the spacecraft can continue operating for a longer period of time, extended missions can be approved with additional science goals. As a spacecraft approaches the end of its operational life (e.g., running out of fuel) an end-of-mission maneuver is designed to safely dispose of the spacecraft. This end-of-mission maneuver can also be used to achieve additional objectives beyond the spacecraft's initial design capabilities. For example, the Cassini spacecraft was disposed of by sending the spacecraft into Saturn's atmosphere,  taking \textit{in situ} measurements that could not be accomplished during the active mission and which resulted in major scientific achievements \citep{Edgington2016}. Additionally, the Rosetta mission ended its life by descending to the surface of its target comet 67P/Churyumov-Gerasimenko and making observations directly at the surface \citep{Accomazzo2017, Barucci2017}. 

At this time, several spacecraft currently in orbit around Mars are approaching their end-of-life \citep{Carr1996}. If the required $\Delta V$ is small, one of these spacecraft could de-orbit from Mars and conduct a flyby of 3I/ATLAS. Such a maneuver would be a potentially high impact end-of-mission scenario for an end-of-life spacecraft. In this section, we assess whether trajectory designs from Mars to 3I/ATLAS may be energetically feasible. Although we consider generic trajectories, our results can be applied to existing spacecraft. We once again use the algorithm discussed in Section \ref{sec:methods} to calculate the minimum-energy trajectories between 3I/ATLAS and Mars.

In Figure \ref{fig:mars_mission_deltav}, we show the minimum $\Delta V$ required for such a mission. The minimum $\Delta V$ of 2.02 km s$^{-1}$ \added{(C$_3$ = 4.08 km$^{2}$ s$^{-2}$)} occurs for a launch on 2025 March 6 (four months prior to discovery), with flight time of 212 days. The first post-discovery opportunity also occurs contemporaneously with discovery and would have required a $\Delta V$ of 3.54 km s$^{-1}$ \added{(C$_3$ = 12.5 km$^{2}$ s$^{-2}$)} on 2025 July 1. A mission with this trajectory would arrive on 2025 October 3, three months after discovery, with its 94 days in transit. This optimal mission would fly by 3I/ATLAS during its closest approach to Mars. In Figure \ref{fig:marsmission_map}, we show the required $\Delta V$ as a function of the Mars departure date and 3I/ATLAS arrival date. In Figure \ref{fig:mars_mission_orbit}, we show the orbit of this minimum-energy post-discovery mission. For detailed information regarding these trajectories, see Table \ref{table:trajectories}.  After discovery, the required $\Delta V$ for flyby climbs gradually, yet remains at $\Delta V\le 10$ km s $^{-1}$ through 2025 August 30.

The sharp rise and gap in $\Delta V$ around October 2025 coincides with the closest approach between Mars and 3I/ATLAS, which occurs on 2025 October 3. \added{Although 3I/ATLAS makes its closest physical approach to Mars shortly after their trajectories cross in the ecliptic (X-Y plane), it is more favorable for a spacecraft to target a flyby near the local node itself. At the node, the orbits are coplanar, so a near-tangent transfer can be achieved with relatively low $\Delta V$. Once 3I/ATLAS passes the node and the separation from Mars increases, achieving a flyby requires substantially higher $\Delta V$, producing the sharp rise in Figure \ref{fig:mars_mission_deltav}.}

The \added{minimum-energy} trajectories sent prior to the closest approach have flight times such that the flyby occurs close in time to 2025 October 3. 
In Figure \ref{fig:mars_mission_orbit_2}, we show the trajectory for a launch on 2025 August 10, 41 days after discovery, which demonstrates one of possible Mars departure scenarios. On the other hand, even the \added{minimum-energy} trajectories after the close approach date would require significantly more $\Delta V$ to perform a flyby.

These results suggest that an intercept or flyby from Mars would require less $\Delta V$ than an Earth-based mission. This advantage is due primarily to orbital phasing. At the perihelion, Mars is closely aligned with 3I/ATLAS orbital longitude, while Earth is nearly $180^\circ$ out of phase, making an Earth‐based intercept far more $\Delta V$ intensive. A hypothetical Mars-based mission to 3I/ATLAS would have $\Delta V$ requirements within the performance envelope of modern spacecraft propulsion systems. Therefore, such a Mars-based flyby mission could be feasible.

\section{Discussion} \label{sec:discussion}

In this paper, we calculated minimum-energy trajectories from both Earth and Mars to the interstellar object 3I/ATLAS. Given the current state of spacecraft propulsion technology, a spacecraft flyby of 3I/ATLAS would be more feasible if launched from Mars rather than Earth.

Our calculations show that post-discovery Earth departures likely demand $\Delta V$ values beyond the capabilities of available spacecraft. However, had 3I/ATLAS been discovered before July 1, a mission launched from Earth would have required $\Delta V\sim7$ km s$^{-1}$, within the envelope of current spacecraft technologies. This fact underscores the critical impact of discovery time on the feasibility of deep-space missions. 

On the other hand, Mars departures require substantially less $\Delta V$. For example, a pre-discovery flyby on 2025 March 6 (four months prior to discovery) could reach 3I/ATLAS with just $\Delta V=2.02$ km s$^{-1}$ while a launch on the discovery date of 2025 July 1 would require $\Delta V\sim4$ km s$^{-1}$. These $\Delta V$ values \added{may} fall within the maneuvering envelope of current Mars orbiters, making it possible to re-target one for a close flyby of 3I/ATLAS. 

There are a variety of spacecraft currently orbiting Mars that are nearing their end of life --- Mars Reconnaissance Orbiter \citep{Zurek2007}, MAVEN \citep{Jakosky2015}, Mars Odyssey \citep{Saunders2004}, Mars Express \citep{Chicarro2004, Chicarro2007}, ExoMars Trace Gas Orbiter \citep{Vandaele2015} and others. These spacecraft collectively carry a wide range of imaging cameras, spectrometers, \textit{in situ} instrumentation, and communications and operation tools. Repurposing these missions as interstellar interceptors would offer valuable scientific data, especially since they may be able to reach interstellar objects that are inaccessible from Earth. While repurposing Earth-observing missions is also possible, these spacecraft are generally de-orbited during their end-of-life phase and have less maneuvering capability, limiting their use as interstellar interceptors. Beyond Mars orbiters, opportunistic geometries with existing deep-space spacecraft may also enable intercepts. \citet{Loeb2025b} propose using \textit{Juno} \citep{Bolton2017} during 3I/ATLAS’s March 2026 Jupiter approach.

The Janus spacecraft mission \citep{Scheeres2021}, part of NASA's SIMPLEx program, consists of two identical \added{small satellites custom-built by Lockheed Martin to be CubeSat-class in scale}, designed to fly by binary asteroids. Each carries JCam, a visible/infrared imager capable of capturing meter-scale resolution images and thermal data. \added{Janus was engineered to perform close flybys at relative velocities of order 10 km s$^{-1}$ and closest approaches of order 50 km for its target binaries. This capability illustrates that compact spacecraft systems can in principle execute and return valuable data from very high-speed encounters with small bodies, similar to what would be required for an interstellar object like 3I/ATLAS.} \added{However, the flyby velocities listed in Table \ref{table:trajectories} for trajectories to 3I/ATLAS all exceed Janus’s demonstrated design regime, highlighting additional technical challenges that would need to be addressed for a mission to 3I/ATLAS.}

Although a flyby may not be feasible, NASA’s OSIRIS-APEX \citep{Dellagiustina2023} could be assigned to observe 3I/ATLAS from the spacecraft's existing trajectory using the PolyCam and MapCam imagers \citep{rizk_ocams_2018}. Observations with the Eight Color Asteroid Survey (ECAS) $b$, $v$, $w$, and $x$ filters onboard the OSIRIS-APEX MapCam instrument \citep{DellaGiustina2018} would provide valuable data characterizing 3I/ATLAS. MapCam could acquire narrow-band photometry from the near-ultraviolet through the near-infrared, constraining the spectral slope and broad surface composition of 3I/ATLAS \citep{Golish2021, Dellagiustina2023}. The viable observation windows are late September and mid-November of 2025. Although late September offers the advantage of pre-perihelion observations, preferable for comet characterization, it coincides with an OSIRIS-APEX Earth Gravity Assist \citep{Dellagiustina2023}, making additional spacecraft activities during this period potentially risky. By mid-November, the solar elongation of 3I/ATLAS increases rapidly, and observations could likely be conducted once the elongation exceeds 35–40 degrees. A spacecraft like OSIRIS-APEX can provide valuable photometric data at a distance, obtaining unique observation geometries in the absence of atmospheric effects.

\added{The OCAMS instruments on OSIRIS-APEX have well-characterized imaging performance. 
PolyCam, with a plate scale of $\sim$13.5 $\mu$rad pixel$^{-1}$ \citep{Golish2020}, 
would achieve a spatial resolution of $\sim$13.5 km pixel$^{-1}$ at 1 million km 
and $\sim$135 km pixel$^{-1}$ at 10 million km. 
MapCam, with a plate scale of $\sim$68 $\mu$rad pixel$^{-1}$, 
would provide $\sim$68 km pixel$^{-1}$ at 1 million km 
and $\sim$680 km pixel$^{-1}$ at 10 million km, while SamCam offers only coarser coverage.}

In addition, higher-resolution imagery could resolve any dust plumes or jets coming off the surface of 3I/ATLAS, further characterizing its mass-loss rate. By viewing the object at different phase angles, the albedo of the object can be constrained (assuming that we can resolve the nucleus). Finally, a simple infrared measurement would indicate how the surface of 3I/ATLAS responds to thermal forcing, enabling further characterization of its surface properties. Together, these results would allow for estimates of nucleus size, composition, rotation, and activity level.

Even if no existing spacecraft performs a flyby of 3I/ATLAS, the discovery and apparition provide a tangible example for a future interstellar object mission. The NSF-DOE Vera C. Rubin Observatory Legacy Survey of Space and Time (LSST) should identify more interstellar objects in the near future \citep{Moro2009,Cook2016,Engelhardt2017,Hoover2022,Marceta2023a,Dorsey2025}. \added{Earlier discovery with surveys would likely have enabled significantly lower $\Delta V$ transfer opportunities for 3I/ATLAS, converting the prohibitive post-discovery trajectories into ones within reach of current launch capabilities.} If we station fueled spacecraft at key locations (e.g., Earth orbit, Mars orbit, the Earth-Sun Lagrange points) we could direct these craft to fly by a newly discovered interstellar object with short notice. If equipped with narrow‑band cameras and sufficient maneuvering capability for small trajectory changes, these fueled spacecraft could execute fast flybys or continuous brightness monitoring of interstellar objects immediately after discovery. Such measurements would provide critical information regarding the composition, rotational state, and activity of interstellar objects beyond the capabilities of ground-based instruments alone. The data provided by such a mission would provide unprecedented insights into the interstellar object population, and in turn, the history of planet formation throughout the Galaxy. 

\section{acknowledgments}

\added{We thank the anonymous reviewers for insightful comments and constructive suggestions that strengthened the scientific content of this manuscript.}

We thank Shannon Curry, Adina Feinstein, Karen Meech, James Wray, Abraham Loeb, Adam Hibberd, Devin Hoover, and Qicheng Zhang for helpful conversations. 

D.Z.S. is supported by an NSF Astronomy and Astrophysics Postdoctoral Fellowship under award AST-2303553.  This research award is partially funded by a generous gift of Charles Simonyi to the NSF Division of Astronomical Sciences. The award is made in recognition of significant contributions to Rubin Observatory’s Legacy Survey of Space and Time. A.Y. and T.F. also acknowledge support from NSF grant number AST-2303553. A.G.T. acknowledges support from the Fannie and John Hertz Foundation and the University of Michigan's Rackham Merit Fellowship Program. E.P.-A. acknowledges support by the Italian Space Agency (ASI) within the LUMIO project (ASI-PoliMi agreement n. 2024-6-HH.0). K.E.M. acknowledges support from the NASA ROSES Discovery Data Analysis Program (DDAP). D.N.D. and M.C.N. acknowledge support under NASA contract NNM10AA11C.

\bibliography{sample7}{}
\bibliographystyle{aasjournal}

\end{document}